\documentclass{ws-ijqi}

\usepackage{epsfig,graphicx,bm}

\newcommand{\beq}{\begin{equation}}
\newcommand{\eeq}{\end{equation}}
\newcommand{\beqa}{\begin{eqnarray}}
\newcommand{\eeqa}{\end{eqnarray}}
\newcommand{\ket} [1] {\vert #1 \rangle}
\newcommand{\bra} [1] {\langle #1 \vert}

\newcommand{\proj}[1]{\ket{#1}\bra{#1}}
\newcommand{\mean}[1]{\langle #1 \rangle}

\begin{document}

\markboth{Nicolas J. Cerf et al.}
{Quantum entanglement enhances the capacity of bosonic channels with memory}

\catchline{}{}{}{}{}

\title{Information transmission via entangled quantum states\\
in Gaussian channels with memory\footnote
{To appear in Proceedings of Workshop on \emph{Quantum entanglement in physical and information sciences},
Pisa, December 14-18, 2004.}}

\author{Nicolas J. Cerf, Julien Clavareau, J\'er\'emie Roland\footnote{Present 
address: Laboratoire de Recherche en Informatique, UMR 8263, 
Universit\'e de Paris-Sud, 91405 Orsay, France}}
\address{QUIC, Ecole Polytechnique, CP 165/59, Universit\'e Libre de Bruxelles, Av. F.D. Roosevelt 50\\
B-1050 Brussels, Belgium}

\author{Chiara Macchiavello}
\address{QUIT, Dipartimento di Fisica ``A. Volta'', Universit\`a di Pavia, 
via Bassi 6\\
I-27100 Pavia, Italy}

\maketitle


\begin{abstract}
Gaussian quantum channels have recently attracted a growing interest, 
since they may lead to a tractable approach 
to the generally hard problem of evaluating quantum channel capacities.
However, the analysis performed so far has always been restricted to 
memoryless channels. 
Here, we consider the case of a bosonic Gaussian channel
with memory, and show that the classical capacity can be significantly 
enhanced by employing entangled input symbols instead of product symbols. 
\end{abstract}

\keywords{Quantum channel capacity; bosonic Gaussian channels}

\section{Introduction}

The  evaluation of information capacities of quantum communication 
channels is of great interest in  quantum information theory. 
In particular, an important
question is to determine how much {\em classical} information can be processed
asymptotically via a quantum channel. This problem has been solved, today,
only for a few quantum channels, and it has been addressed only recently
for bosonic channels, i.e., continuous-variable quantum channels acting
on a bosonic field such as the electromagnetic field \cite{holevo-werner}.
The classical capacity of a purely {\em lossy} bosonic channel was solved exactly very recently \cite{lossy-capacity}, while  
the case of {\em noisy} bosonic channels is already more involved. Actually,
the classical capacity of the Gaussian bosonic channel, i.e., 
a continuous-variable quantum channel undergoing a Gaussian-distributed 
thermal noise, has been derived in \cite{holevoetal}
although this result only holds provided that the optimal input ensemble 
is a tensor product of Gaussian states, as conjectured by several authors
but not rigorously proven today (see e.g. \cite{giovannetti} for recent progress
on this problem). All these studies, however, have been restricted to memoryless bosonic channels.
\par

The first analysis of the classical capacity of quantum channels with memory 
was performed in \cite{macch-palma}.
In this work a qubit depolarizing channel with memory was investigated and 
the first evidence that entangled input states enhance the single-shot 
capacity was provided. It was shown that the single-shot capacity is always 
maximized either by product states or by maximally entangled states at the input,
and that the product versus entangled states optimal regime depends on
a threshold value for the degree of channel memory. This threshold value
turned out to be a function  of the depolarization factor of the channel.
The same features mentioned above for the depolarizing 
channel with memory were also later proved for a class of non-isotropic qubit 
channels with memory \cite{mpv}.

In this paper, we investigate the capacity of a bosonic Gaussian channel 
that exhibits memory. We review the main results that were first presented in \cite{PRA} and
provide more details about their derivation.
We consider channels with a thermal noise that has a {\em finite} bandwidth.
The resulting memory effect is modeled by assuming that the noise
affecting two subsequent uses of the channel follows a bivariate Gaussian
distribution with a non-vanishing correlation coefficient, measuring the
degree of memory of the channel. We prove that if the memory
is non-zero and if the input energy is constrained, 
then the channel capacity can be significantly
enhanced by using entangled symbols instead of product symbols. 
The relation between the degree of memory and the resulting optimal input entanglement is analyzed.

The paper is organized as follows. In Sect. 2 we review the description and
known results about memoryless bosonic Gaussian channels.
In Sect. 3 we introduce the role of memory, and analyse the capacity 
by allowing the transmission of entangled input states. Finally, in Sect. 4
we summarise and discuss our results.

\section{Memoryless Bosonic Gaussian channels}
Let us define a memoryless bosonic Gaussian channel $T$ 
acting on a mode of the electromagnetic field associated with
the annihilation and creation operators $a$ and $a^\dagger$,
or, equivalently, the quadrature components $q=(a+a^\dagger)/\sqrt{2}$
and $p=i(a^\dagger-a)/\sqrt{2}$, satisfying
the commutation relation $[q,p]=i$.
If the input of the channel is initially in state $\rho$, we have
\beq
\rho\mapsto T[\rho]=\int d^2\beta\ q(\beta)\ D(\beta)\rho D^\dagger(\beta),
\eeq
where $d^2\beta=d\Re(\beta) \, d\Im(\beta)$, 
while $D(\beta)=e^{\beta a^\dagger-\beta^* a}$ denotes 
the displacement operator (such that $\ket{\alpha}=D(\alpha)\ket{0}$
with $\ket{0}$ being the vacuum state and $\ket{\alpha}$ being a coherent 
state of mean value $\alpha$). For a Gaussian channel, the kernel
is a bivariate Gaussian distribution with variance $N$, namely
\beq
q(\beta)=\frac{1}{\pi N} \, e^{-\frac{|\beta|^2}{N}}\;.
\eeq
The channel then randomly displaces an input coherent state
according to a Gaussian distribution, which results 
in a thermal state ($N$ is the variance of the added noise
on the quadrature components $q$ and $p$, or, equivalently
the number of thermal photons added by the channel). 
The Gaussian CP map effected by this channel
can also be characterized via the covariance matrix.
Restricting to Gaussian states with a vanishing mean value, 
a complete state characterization is provided by the covariance matrix
\beqa
\gamma&=&\left(\begin{array}{cc}
\mean{q^2} & \frac{1}{2}\mean{qp+pq} \\ 
\frac{1}{2}\mean{qp+pq} & \mean{p^2}
\end{array} \right).
\eeqa
The Gaussian channel can then be written as
\beq
\gamma \mapsto \gamma + \left(\begin{array}{cc}
N & 0 \\ 
0 & N
\end{array} \right).
\eeq

The coding theorem for quantum channels asserts 
that the one-shot classical capacity of a quantum
channel $T$ is given by
\beq\label{capacity}
C_1(T)=\max\left[S\left(\sum_i p_i T[\rho_i]\right)-\sum_i p_i S\left(T[\rho_i]\right)\right],
\eeq
where 
$S(\rho)=-\textrm{Tr}(\rho\log\rho)$
is the von Neumann entropy of the density operator $\rho$.
In Eq.~(\ref{capacity}),
the maximum is taken over all probability distributions $\{p_i\}$ and collections of density operators $\{\rho_i\}$
satisfying the energy constraint
\beq
\sum_i p_i \textrm{Tr}\left(\rho_i a^\dagger a\right)\leq\bar{n},
\eeq
with $\bar{n}$ being the maximum mean photon number at the input
of the channel. For a monomodal bosonic Gaussian channel,
it is conjectured that a Gaussian mixture of coherent states
(i.e., a thermal state) achieves the channel capacity \cite{fn1}.
The sum over $i$ is replaced by an integral over $\alpha$,
where the input states
\beq
\rho_\alpha^{\textrm{in}}=\proj{\alpha}
\eeq
are
drawn from the probability density
\beq
p(\alpha)=\frac{1}{\pi \bar{n}} \, e^{-\frac{|\alpha|^2}{\bar{n}}}.
\eeq
Thus, the one-shot classical capacity of the channel becomes
\beq
C_1(T)=S\left(\bar{\rho}\right)-\int d^2\alpha \,  p(\alpha) \, S\left(\rho_\alpha^{\textrm{out}}\right),
\eeq
where we have defined the individual output states
\beq
\rho_\alpha^{\textrm{out}}=T[\rho_\alpha^{\textrm{in}}]=\frac{1}{\pi N}\int d^2\beta \, e^{-\frac{|\beta-\alpha|^2}{N}} \, \proj{\beta}
\eeq
and their mixture (saturating the energy constraint)
\beq
\hspace*{-2mm}\bar{\rho}=\int d^2\alpha \, p(\alpha) \, \rho_\alpha^{\textrm{out}}
=\frac{1}{\pi(\bar{n}+N)}\int d^2\beta \, e^{-\frac{|\beta|^2}{\bar{n}+N}} \, \proj{\beta}.
\eeq
In order to calculate the entropy of a Gaussian state $\rho$, 
one computes the symplectic values of its covariance matrix
$\gamma$, i.e., the solutions 
of the equation $\left|\gamma-\lambda J\right|=0$,
where
\beq\label{matrix-j}
J=\left(\begin{array}{cc}
0 & i \\ 
-i & 0
\end{array} \right).
\eeq
It can be shown that these values always come as a pair $\pm\lambda$, so that
the entropy is given by $S(\rho)=g\left(|\lambda|-\frac{1}{2}\right)$,
where
\beq
g(x)= \left\{ \begin{array}{ll}
(x+1)\log_2 (x+1)-x\log_2 x,\qquad & x>0\\
0\qquad & x=0 \end{array} \right.
\eeq
is the entropy of a thermal state with a mean photon number of $x$.
Since the input states $\rho_\alpha^{\textrm{in}}$ are coherent states with a covariance matrix
\beq\label{matrix-coherent}
\gamma^\textrm{in}=\frac{1}{2}
\left(\begin{array}{cc}
1 & 0 \\ 
0 & 1
\end{array} \right),
\eeq
the individual output states $\rho_\alpha^\textrm{out}$ and their mixture $\bar{\rho}$ are associated with the covariance matrices
\beqa
\gamma^\textrm{out}&=&\frac{1}{2}
\left(\begin{array}{cc}
1+2N & 0 \\ 
0 & 1+2N
\end{array} \right),\\
\bar{\gamma}&=&\frac{1}{2}
\left(\begin{array}{cc}
1+2(\bar{n}+N) & 0 \\ 
0 & 1+2(\bar{n}+N)
\end{array} \right),
\eeqa
so that the one-shot capacity of the channel is
\beq
C_1(T)=g(\bar{n}+N)-g(N).
\eeq

\section{Bosonic Gaussian channels with memory}
\subsection{Multimodal Gaussian states}
Let us now generalize these notions to multimodal channels.
Similarly to the monomodal case, $s$ modes of the electromagnetic field, numbered from $1$ to $s$, are associated to $s$ pairs of annihilation and creation
operators $a_j$ and $a^\dagger_j$ or, equivalently, of quadrature components  $q_j$ and $p_j$.
Ordering these observables in a column vector
\beq\label{col-vector}
R=[q_1,p_1,\ldots,q_s,p_s]^\textrm{T},
\eeq
we define the mean vector $m^{(s)}$ and the covariance matrix $\gamma^{(s)}$ of an $s$-mode state $\rho^{(s)}$ as
\beqa
m^{(s)}&=&\textrm{Tr}\ \rho^{(s)} R,\\
\gamma^{(s)}&=&\textrm{Tr}\ (R-m^{(s)})\rho(R-m^{(s)})^\textrm{T}-\frac{1}{2}\bigoplus_{j=1}^s J_j,\label{cov-matrix}
\eeqa
where each $J_j$ takes the form (\ref{matrix-j}). We now focus on Gaussian states, which are completely specified
by their mean vector and covariance matrix. The von Neuman entropy of such a state is then given by
\beq
S(\rho^{(s)})=\sum_{j=1}^s g\left(|\lambda_j|-\frac{1}{2}\right),
\eeq
where $\pm\lambda_j$ are the symplectic eigenvalues of the covariance matrix $\gamma^{(s)}$ of the state; that is the solutions of the equation
\beq
\left|\gamma^{(s)}-\lambda \bigoplus_{j=1}^s J_j\right|=0.
\eeq
This will allow us to compute the one-shot classical capacity of bosonic Gaussian channels acting on $s$ modes \cite{fn1},
since this will still take the form (\ref{capacity}), but with a generalized energy constraint
\beq
\frac{1}{s}\sum_{j=1}^s\sum_{i} p_i \textrm{Tr}\left(\rho_i^{(s)} a_j^\dagger a_j\right)\leq\bar{n}.
\eeq

\subsection{Memoryless bimodal channel}
Consider two subsequent uses of a memoryless channel $T$, 
defining the bimodal channel 
\beqa
\lefteqn{\rho \mapsto T_{12}[\rho]=\int d^2\beta_1 \ d^2\beta_2 \  q(\beta_1,\beta_2)} \hspace{0.5cm} \nonumber \\
&\times& D(\beta_1)\otimes D(\beta_2)\; \rho \; D^\dagger(\beta_1)\otimes D^\dagger(\beta_2),
\eeqa
where
\beq
q(\beta_1,\beta_2)=\frac{1}{\pi^2 N^2} e^{-\frac{|\beta_1|^2+|\beta_2|^2}{N}}
\eeq
since the noise affecting the two uses is uncorrelated. Following Eqs.~(\ref{col-vector}-\ref{cov-matrix}), we 
define the covariance matrix $\gamma_{12}$ of a bimodal state $\rho_{12}$.
We restrict ourselves to bimodal Gaussian states \cite{fn1},
characterized by 
\beq\label{generic-covariance}
\gamma_{12}=\left(\begin{array}{cc}
\gamma_1 & \sigma_{12} \\ 
\sigma_{12}^\mathrm{T} & \gamma_2
\end{array}\right),
\eeq
where $\gamma_1$ is the covariance matrix associated
with the reduced density operator $\rho_1=\textrm{Tr}_2(\rho_{12})$ 
of mode $1$ (and similarly for $\gamma_2$),
while $\sigma_{12}$ characterizes the correlation and/or entanglement 
between the two modes. For a memoryless channel,
the optimal input states are simply products of coherent states, 
with a covariance matrix
$\gamma_{12}^\textrm{in}=\gamma_1^\textrm{in}\oplus\gamma_2^\textrm{in}$
where $\gamma_1^\textrm{in}$ and $\gamma_2^\textrm{in}$ both take the form (\ref{matrix-coherent}), while $\sigma_{12}^\textrm{in}=0$. 
The optimal input modulation is a product of Gaussian distributions,
\beq
p(\alpha_1,\alpha_2) = \frac{1}{\pi^2 \bar{n}^2} e^{-\frac{|\alpha_1|^2+|\alpha_2|^2}{\bar{n}}}.
\eeq
It follows that the classical capacity of this channel is additive
$\frac{1}{2} \times C_1(T_{12})=C_1(T)$.

\subsection{Bimodal channel with correlated noise}
Let us investigate what happens
if the noise is correlated, for instance when the two uses 
are closely separated in time and the channel has a finite bandwidth.
We assume that the noise distribution
takes the general form
\beq
q(\beta_1,\beta_2)=\frac{1}{\pi^2 \sqrt{|\gamma^N|}}
e^{-\bm{\beta}^\dagger{\gamma^N}^{-1}\bm{\beta}},
\eeq
where 
$\bm{\beta}=[\Re(\beta_1),\Im(\beta_1),\Re(\beta_2),\Im(\beta_2)]^\textrm{T}$
and $\gamma^N$ is the covariance matrix of the noise quadratures,
chosen to be
\beq
\gamma^N=\left(\begin{array}{cccc}
N & 0 & -xN & 0 \\ 
0 & N & 0 & xN \\ 
-xN & 0 & N & 0 \\ 
0 & xN & 0 & N
\end{array} \right).
\eeq
Thus, the map $T_{12}$ can be expressed by
$\gamma_{12} \mapsto \gamma_{12} + \gamma^N$,
so that the noise terms added on the $p$ quadratures 
of modes 1 and 2 are correlated Gaussians with variance $N$, while those added
on the $q$ quadratures are anticorrelated Gaussians with variance $N$.
In Sect.~\ref{sec-phase}, we will see that the beneficial effect of entanglement disappears
in a more symmetric noise model where the $q$ and $p$ noise quadratures
are both correlated.
The correlation coefficient $x$ ranges from
$x=0$ for a memoryless channel to $x=1$ for a channel
with full memory.
\par

We now come to the central result of this paper. While we have seen that
for a memoryless channel, the capacity is
attained for product states, we will prove that for correlated thermal noise,
the capacity is achieved if some appropriate degree of entanglement
is injected at the input of the channel. 
Intuitively, if we take an EPR state, i.e., the common eigenstate
of $q_+=(q_1+q_2)/\sqrt{2}$ and $p_-=(p_1-p_2)/\sqrt{2}$,
it is clear that the noise on $q_+$ and $p_-$ effected by the channel
is reduced as $x$ increases. This suggests that using entangled
input states may decrease the effective noise, hence increase the capacity.
However, EPR states have infinite energy so they violate
the energy constraint.
Instead, we may inject (finite-energy) two-mode vacuum squeezed states,
whose covariance matrix is given by
\beqa
\gamma_1^\textrm{in}&=&\gamma_2^\textrm{in}=\frac{1}{2}\left(
\begin{array}{cc}
\cosh 2r & 0 \\ 
0 & \cosh 2r
\end{array} 
\right),   \label{EPR1}\\
\sigma_{12}^\textrm{in}&=&\frac{1}{2}\left(
\begin{array}{cc}
-\sinh 2r & 0 \\ 
0 & \sinh 2r
\end{array} 
\right),  \label{EPR2}
\eeqa
with $r$ being the squeezing parameter.
Note that purely classical correlations between 
the quadratures in the input distribution $p(\alpha_1,\alpha_2)$
also help increase the capacity when $x>0$, so we have to check
that entanglement gives an extra enhancement in addition to this.
\par

The mean photon number in each mode of the state characterized by 
Eqs.~(\ref{EPR1})-(\ref{EPR2}) is $\sinh^2 r$, 
so that the maximum allowed modulation (for a fixed maximum photon number
$\bar{n}$) decreases as entanglement increases. Remarkably, there 
is a possible compromise between this reduction of modulation 
and the entanglement-induced noise reduction on $q_+$ and $p_-$. 
To show this, consider input states with $\sinh^2 r = \eta \bar{n}$; that is
\beqa
\gamma_{1,2}^\textrm{in}&=&\frac{1}{2}\left(
\begin{array}{cc}
1+2\eta\bar{n} & 0 \\ 
0 & 1+2\eta\bar{n}
\end{array} 
\right),\label{gammain}\\
\sigma_{12}^\textrm{in}&=&\frac{1}{2}\left(
\begin{array}{cc}
-2\sqrt{\eta\bar{n}(1+\eta\bar{n})} & 0 \\ 
0 & 2\sqrt{\eta\bar{n}(1+\eta\bar{n})}
\end{array} \label{sigmain}
\right),
\eeqa
where $\eta$ measures the {\em degree of entanglement} 
and is used to interpolate between
a product of vacuum states ($\eta=0$), which can be maximally modulated,
and an entangled state ($\eta=1$), for which the entire
energy is due to entanglement and no modulation can be applied.
At the output of the channel, we get 
states with a covariance matrix $\gamma_{12}^\textrm{out}$ where
\beqa
\gamma_{1,2}^\textrm{out}&=&\frac{1}{2}\left(
\begin{array}{cc}
1+2\eta\bar{n}+2N & 0 \\ 
0 & 1+2\eta\bar{n}+2N
\end{array} 
\right),   \label{gammaout}\\
\sigma_{12}^\textrm{out}&=&\frac{1}{2}\left(
\begin{array}{cc}
-2\sqrt{\eta\bar{n}(1+\bar{n})}-2xN & 0 \\ 
0 & 2\sqrt{\eta\bar{n}(1+\bar{n})}+2xN
\end{array} 
\right),
\eeqa
while the mixture $\bar{\gamma}_{12}$ of these states are characterized by
\beqa
\bar{\gamma}_{1,2}&=&
\gamma_{1,2}^\textrm{out}+\left(
\begin{array}{cc}
(1-\eta)\bar{n} & 0 \\ 
0 & (1-\eta)\bar{n}
\end{array} 
\right),\\
\bar{\sigma}_{12}&=& \sigma_{12}^\textrm{out} +\left(
\begin{array}{cc}
y (1-\eta)\bar{n} & 0 \\ 
0 & -y (1-\eta)\bar{n}
\end{array} 
\right), \label{sigmamean}
\eeqa
assuming that the energy constraint is saturated. Here,
$y$ stands for the classical input correlation coefficient
(to compensate for the noise,
the $q$ displacements need to be correlated, 
and the $p$ displacements anti-correlated).

\subsection{Independent channels with phase-sensitive noise\label{sec-phase}}
Before evaluating the transmission rate achieved by these states, let us give some
intuition about why entangled states could help enhance the transmission rate.
More precisely, let us study this channel in a modified basis by
considering the transformation
\beq
\left\{\begin{array}{ccc}
q_+&=&{(q_1+q_2)}/{\sqrt{2}}\\
p_+&=&{(p_1+p_2)}/{\sqrt{2}}
\end{array}\right.
\quad\textrm{and}\quad
\left\{\begin{array}{ccc}
q_-&=&{(q_1-q_2)}/{\sqrt{2}}\\
p_-&=&{(p_1-p_2)}/{\sqrt{2}}
\end{array}\right.
\eeq
which physically corresponds to a 50/50 beam splitter and maps the $1$ and $2$ modes
to the $+$ and $-$ modes. Expressing the covariance matrix $\gamma_{+-}^N$ of the noise
in this new basis confirms that the noise on the $q_+$ and $p_-$ quadratures
is reduced, while the noise on the $p_+$ and $q_-$ is enhanced:
\beq
\gamma_{+-}^N=\left(\begin{array}{cccc}
N(1-x) & 0 & 0 & 0 \\ 
0 & N(1+x) & 0 & 0 \\ 
0 & 0 & N(1+x) & 0 \\ 
0 & 0 & 0 & N(1-x)
\end{array} \right).
\eeq
Moreover, in this basis, the channel introduces no correlation between the $+$ and $-$ modes,
so that it acts as a product of two \emph{independent} (memoryless) monomodal channels.
These monomodal channels are \emph{phase-sensitive}, in contrast to those acting on modes $1$ and $2$,
since the variances of the $q$ and $p$ noise
quadratures are different. As a consequence, their capacity is achieved
by transmitting squeezed states \cite{holevoetal},
described by covariance matrices
\beqa
\gamma_+^\textrm{in}&=&
\frac{1}{2}\left(
\begin{array}{cc}
e^{-2r} & 0 \\ 
0 & e^{2r}
\end{array} 
\right),\\
\gamma_-^\textrm{in}&=&
\frac{1}{2}\left(
\begin{array}{cc}
e^{2r} & 0 \\ 
0 & e^{-2r}
\end{array} 
\right),
\eeqa
where the squeezing parameter $r$ is, as above, such that $\sinh^2 r = \eta \bar{n}$.
Rotating back this product of squeezed states, corresponding to a diagonal covariance matrix
$\gamma_{+-}^\textrm{in}=\gamma_+^\textrm{in}\oplus\gamma_-^\textrm{in}$,
into the original basis, we obtain the two-mode squeezed states
with covariance matrix $\gamma_{12}^\textrm{in}$ as defined in Eqs.~(\ref{gammain})-(\ref{sigmain}).

We may also continue studying this channel in the modified basis where all covariance
matrices are diagonal. Indeed, the covariance matrix of states at the output of the channel
is given by $\gamma_{+-}^\textrm{out}=\gamma_{+-}^\textrm{in}+\gamma_{+-}^N$.
As for the mixture of output states, we have used $\eta\bar{n}$ photons for squeezing,
so that we are left with $(1-\eta)\bar{n}$ photons for modulation. Distributing this photon number between
the quadratures following the affecting noise then leads to a covariance matrix
\beq
\bar{\gamma}_{+-}=\gamma_{+-}^\textrm{out}+\left(\begin{array}{cccc}
(1+y)(1-\eta)\bar{n} & 0 & 0 & 0 \\ 
0 & (1-y)(1-\eta)\bar{n} & 0 & 0 \\ 
0 & 0 & (1-y)(1-\eta)\bar{n} & 0 \\ 
0 & 0 & 0 & (1+y)(1-\eta)\bar{n}
\end{array} \right),
\eeq
where the parameter $y$ allows to optimize the distribution of the modulation between the quadratures.
Up to the basis change, $\bar{\gamma}_{+-}$ is equivalent to the covariance matrix $\bar{\gamma}_{12}$
characterized by Eqs.~(\ref{gammaout})-(\ref{sigmamean}).

Using this basis, we also easily understand why the transmission rate of
a bosonic gaussian channel where the $q$ and $p$
noise quadratures are both correlated (or both anti-correlated) may not be enhanced by using
entangled states. Indeed, the covariance matrix of the noise on the $+$ and $-$ modes
takes in this case the form
\beq
\gamma_{+-}^N=\left(\begin{array}{cccc}
N(1-x) & 0 & 0 & 0 \\ 
0 & N(1-x) & 0 & 0 \\ 
0 & 0 & N(1+x) & 0 \\ 
0 & 0 & 0 & N(1+x)
\end{array} \right),
\eeq
which is characteristic of a product of two independent \emph{phase-insensitive} monomodal channels since
the variance of the noise affecting both quadratures of a given mode is the same. The capacity
of this channel is thus achieved by transmitting coherent states, which, after the change of basis,
remain unentangled states.

\subsection{Entanglement-enhanced capacity}
In order to evaluate the transmission rate achieved by these states, we need
first to compute the symplectic values $\lambda_{12}^\textrm{out}$
and $\bar{\lambda}_{12}$ of $\gamma_{12}^\textrm{out}$
and $\bar{\gamma}_{12}$, respectively (or, equivalently, of $\gamma_{+-}^\textrm{out}$
and $\bar{\gamma}_{+-}$).
The symplectic values $\pm\lambda_{12}$ of a covariance matrix $\gamma_{12}$ of the generic form
(\ref{generic-covariance}) are the solutions of the equation
\beq
\left|\gamma_{12}-\lambda_{12} (J_1\oplus J_2)\right|=0\;,
\eeq
or, equivalently, the biquadratic equation
\beq
\lambda_{12}^4-(|\gamma_1|+|\gamma_2|+2|\sigma_{12}|)\lambda_{12}^2
+|\gamma_{12}|=0.
\eeq
Using Eqs.~(\ref{gammaout})-(\ref{sigmamean}), we see that
$\gamma_{12}^\textrm{out}$ and $\bar{\gamma}_{12}$ admit each one pair of doubly-degenerate symplectic values, namely
\beq
\lambda_{12}^\textrm{out}=\pm \sqrt{u_\textrm{out}^2-v_\textrm{out}^2}\;, 
\qquad
\bar{\lambda}_{12}=\pm \sqrt{\bar{u}^2-\bar{v}^2}\;, 
\eeq
with
\beqa
u_\textrm{out}= \textstyle{\frac{1}{2}}+\eta\bar{n}+N, \hspace{1.2cm}
v_\textrm{out}=\sqrt{\eta\bar{n}(1+\eta\bar{n})}+xN , \nonumber\\
\bar{u}=\textstyle{\frac{1}{2}}+\bar{n}+N, \quad
\bar{v}=\sqrt{\eta\bar{n}(1+\eta\bar{n})}+xN-y(1-\eta)\bar{n}.\nonumber
\eeqa
The transmission rate per mode is then given by
\beq
R(\eta,y)
=g(|\bar{\lambda}_{12}|-\textstyle{\frac{1}{2}})-g(|\lambda_{12}^\textrm{out}|-\textstyle{\frac{1}{2}})
\eeq

\begin{figure}[t]
\begin{center}
\epsfig{figure=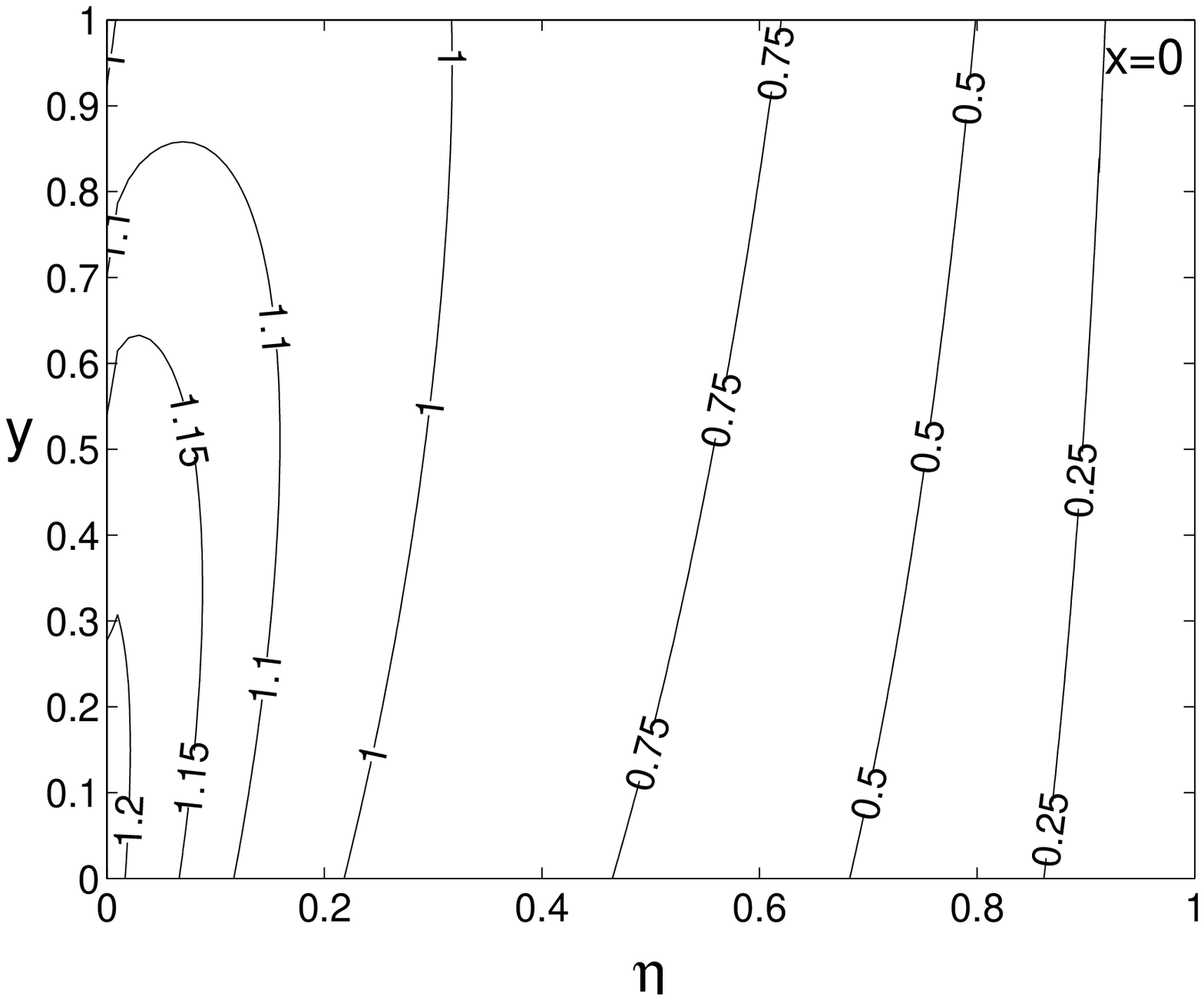,width=50mm}\hspace{5mm}
\epsfig{figure=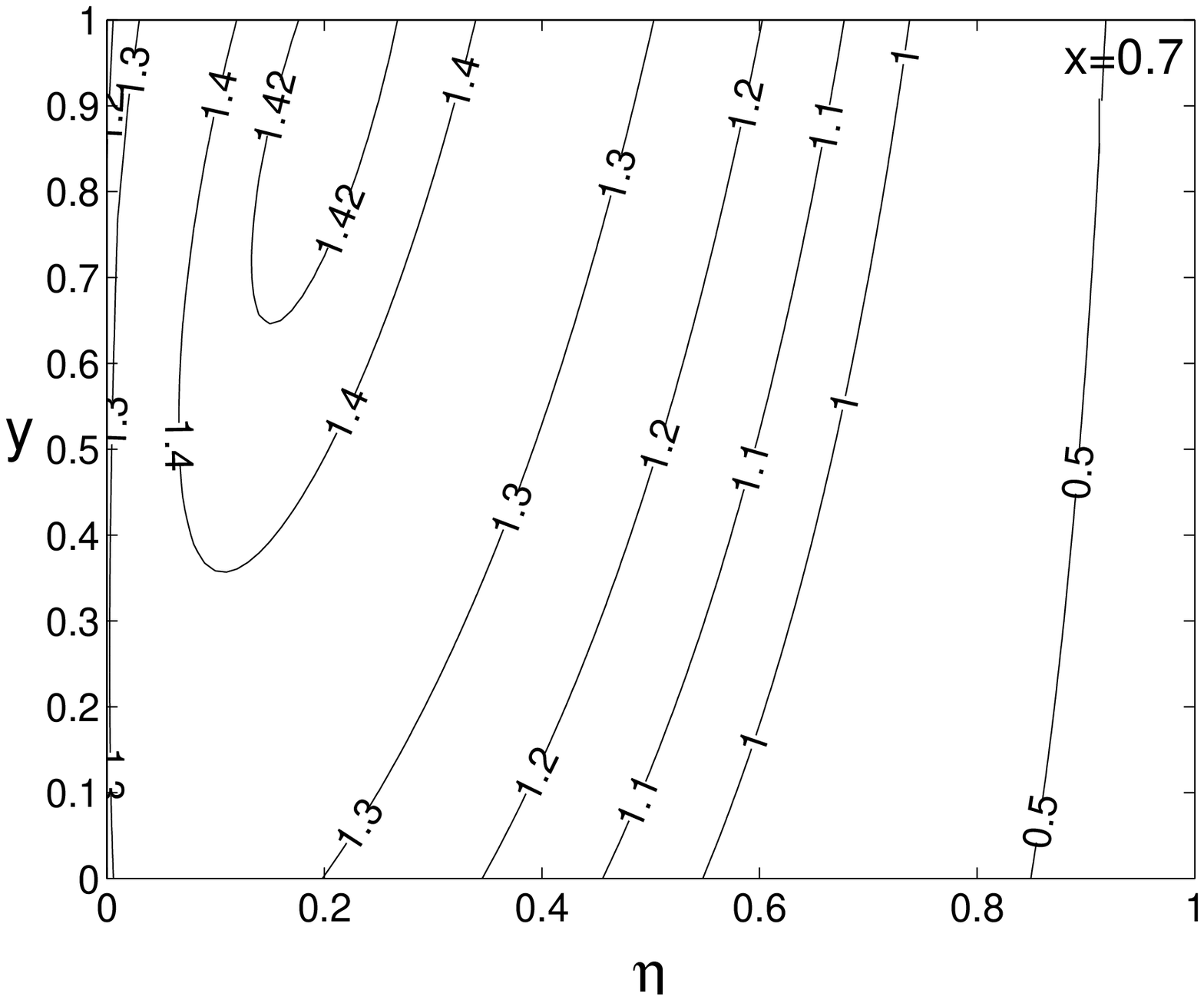,width=50mm}\\
\vspace{5mm}
\epsfig{figure=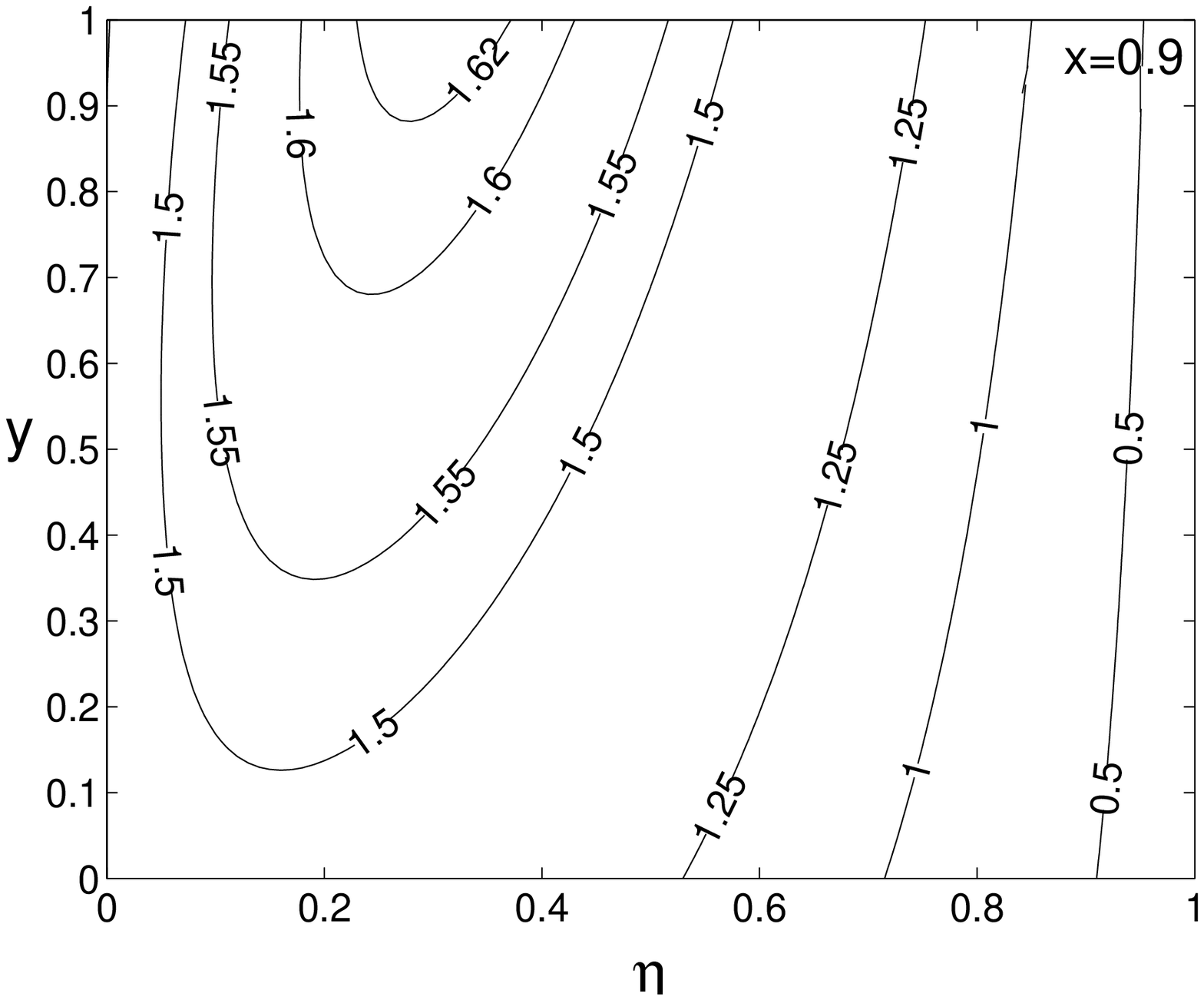,width=50mm}\hspace{5mm}
\epsfig{figure=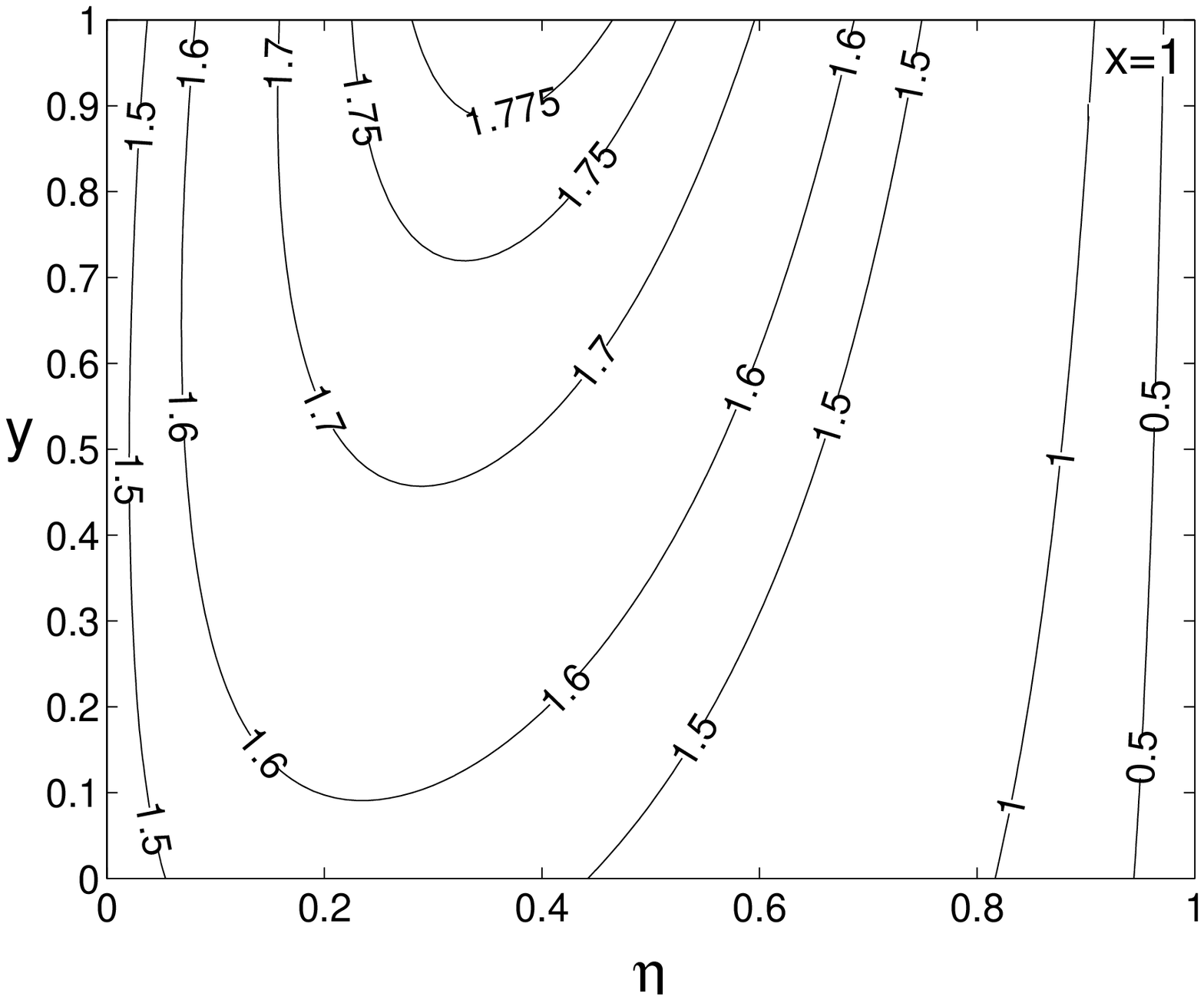,width=50mm}
\caption{Iso-transmission rate lines in the $(\eta,y)$-plane, where $\eta$ is the degree of entanglement
and $y$ is the classical
correlation coefficient, for a thermal channel with a degree of memory $x=0$, $0.7$, $0.9$ and $1$. The mean
number of photons is $\bar{n}=1$ at the input, while the number of added
thermal photons is $N=1/3$.}
\label{fig0}
\end{center}
\end{figure}

We may now vary the degree of entanglement $\eta$ and the classical correlation $y$ to maximize
the transmission rate $R(\eta,y)$. In Fig.~\ref{fig0}, where the level curves of $R(\eta,y)$
for $\bar{n}=1$ and $N=1/3$ are plotted, we see that when there is no memory ($x=0$),
the maximum lies at the origin $(\eta_*,y_*)=(0,0)$, while the transmission rate decreases as soon
as we add classical correlation or entanglement. However, when $x>0$, the
maximum is shifted towards non-zero $\eta$ and $y$ values, that is,
correlation \emph{and} entanglement helps increase the transmission rate.
For instance, when $x=0.7$, the maximum is attained for $(\eta_*,y_*)=(0.19,0.88)$.
When the degree of memory $x$ is further increased, the optimal transmission rate is achieved
by using maximal classical correlation $y=1$, e.~g., for $x=0.9$, we have $(\eta_*,y_*)=(0.30,1)$.
Nonetheless, as $x$ increases towards $x=1$,
the transmission rate may still be enhanced by increasing the degree of entanglement $\eta$.
At $x=1$, we have $(\eta_*,y_*)=(0.37,1)$.

\begin{figure}[t]
\begin{center}
\epsfig{figure=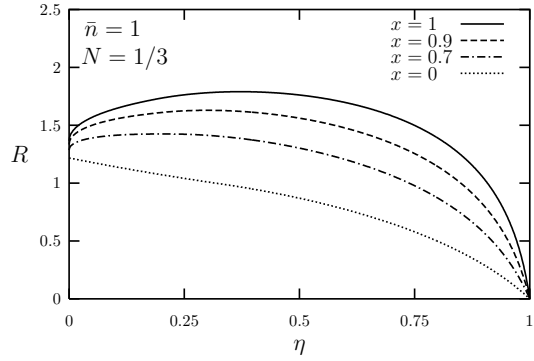,width=70mm}
\caption{Transmission rate $R$ (maximized over the classical
correlation coefficient $y$)
as a function of the input entanglement $\eta$
for a thermal channel with a degree of memory $x$. The mean
number of photons is $\bar{n}=1$ at the input, while the number of added
thermal photons is $N=1/3$.}
\label{fig1}
\end{center}
\end{figure}

In Fig.~\ref{fig1}, we have plotted the transmission rate $R$ optimized
over the classical correlation $y$, as a function of the degree of entanglement $\eta$.
When $x > 0$, the optimized rate $R$ over $y$ increases with the degree of entanglement $\eta$
and attains a maximum at some optimal value $\eta^* > 0$.
This confirms that entangled input states are useful to enhance the transmission rate provided
that the channel exhibits some memory.
\par

\begin{figure}[t]
\begin{center}
\epsfig{figure=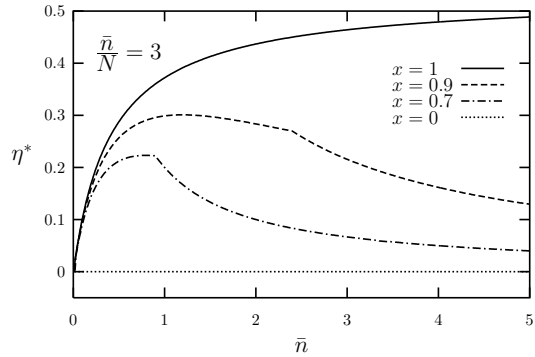,width=70mm}
\caption{Optimal degree of input entanglement $\eta^*$ as a function
of the mean input photon number $\bar{n}$ for a fixed signal-to-noise
ratio $\bar{n}/N=3$.}
\label{fig2}
\end{center}
\end{figure}

\begin{figure}[t]
\begin{center}
\epsfig{figure=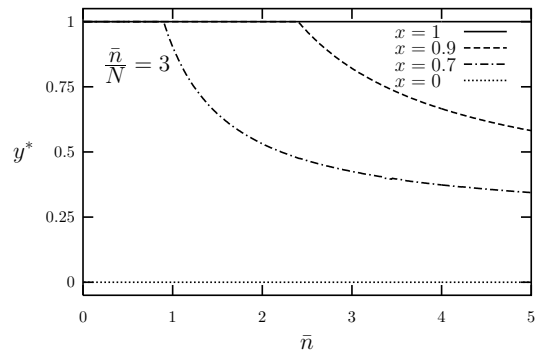,width=70mm}
\caption{Optimal degree of input correlation $y^*$ as a function
of the mean input photon number $\bar{n}$ for a fixed signal-to-noise ratio $\bar{n}/N=3$.}
\label{fig3}
\end{center}
\end{figure}

It now suffices to maximize $R$ with respect to both $y$ and $\eta$ 
in order to find the channel capacity $C$ (assuming that the conjecture \cite{fn1} is verified and that no product but non-Gaussian states may outperform the Gaussian entangled states considered here).
If we keep the signal-to-noise
ratio $\bar{n}/N$ constant, it is visible from Fig.~\ref{fig2} that
the optimal degree of entanglement $\eta^*$ is the highest 
at some particular value of the mean input photon number $\bar{n}$, 
and then decreases back to zero
in the large-$\bar{n}$ limit (except if $x=0$ or $1$).
Clearly, in this limit, the channel $T$ tends to a couple of 
classical channels with Gaussian additive noise (one for each quadrature),
so that entanglement cannot play a role any more \cite{fn2}. 
Fig.~\ref{fig3} shows the corresponding optimal value
of the input correlation coefficient $y^*$ 
for the same values of the other relevant parameters. Note that,
even in the classical limit $\bar{n}\to\infty$,
some non-zero input correlation is useful to enhance the capacity 
of a Gaussian channel with $x>0$.

\section{Conclusion}
\begin{figure}[h]
\begin{center}
\epsfig{figure=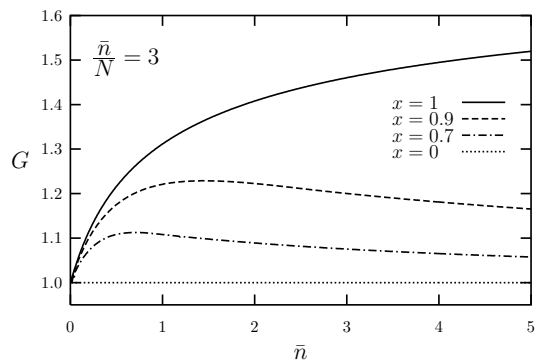,width=70mm}
\caption{Capacity gain $G=\max_{y,\eta}R(\eta,y)/\max_y R(0,y)$
as a function of the mean input photon number $\bar{n}$
for a fixed signal-to-noise ratio $\bar{n}/N=3$.}
\label{fig4}
\end{center}
\end{figure}

We have shown that entangled states can be used to enhance
the classical capacity of a bosonic channel undergoing a thermal noise with
memory. We determined the amount of entanglement that maximizes 
the information transmitted over the channel for a given input energy constraint
(mean photon number per mode) and a given noise level (mean number of thermal photons per mode). For example, the capacity of a channel with a mean number 
of thermal photons of 1/3 and a correlation coefficient of 70\% is enhanced
by 10.8\% if the mean photon number is 1 and the two-mode squeezing is 3.8 dB
at the input. This capacity enhancement may seem paradoxical 
at first sight since using entangled signal states necessarily
decreases the modulation variance for a fixed input energy, which
seemingly lowers the capacity. However, due to the quantum correlations
of entangled states, the noise affecting one mode can be partly
compensated by the correlated noise affecting the second mode, 
which globally reduces the effective noise. 
Interestingly, there exists a regime in which this latter effect dominates,
resulting in a net enhancement of the amount of classical information
transmitted per use of the channel.
The capacity gain $G$, measuring the entanglement-induced capacity enhancement, 
is plotted in Fig.~\ref{fig4}. It illustrates that a capacity enhancement 
of tens of percents is achievable by using entangled light beams with
experimentally accessible levels of squeezing.
It is interesting to notice that, unlike the qubit channels 
investigated so far, in the continuous variable case the capacity is always
optimized by entangled states by varying the degree of entanglement, and that
no threshold on the degree of memory is present.

\section*{Acknowledgments}
We thank V. Giovannetti for informing us of his recent
related work on bosonic memory channels, see e-print quant-ph/0410176.
We are also very grateful to A. Holevo and an anonymous referee of \cite{PRA} for useful comments.
We acknowledge financial support from the Communaut\'e Fran\c caise 
de Belgique under grant ARC 00/05-251, from the IUAP programme of the Belgian government under grant V-18, and from the EU under projects COVAQIAL and
QUPRODIS.
J.R. acknowledges support from the Belgian FRIA foundation.


\begin{thebibliography}{99}
\bibitem{holevo-werner} A. S. Holevo and R. F. Werner,
 Phys. Rev. A \textbf{63}, 032312 (2001).
\bibitem{lossy-capacity} V. Giovannetti, S. Guha, S. Lloyd, L. Maccone,
 J. H. Shapiro, and H. P. Yuen, Phys. Rev. Lett. \textbf{92}, 027902 (2004).
\bibitem{holevoetal} A. S. Holevo, M. Sohma, and O. Hirota, Phys. Rev. A
  \textbf{59}, 1820 (1999).
\bibitem{giovannetti} V. Giovannetti, S. Lloyd, L. Maccone, J. H. Shapiro,
and B. J. Yen, Phys. Rev. A \textbf{70}, 022328 (2004);
V. Giovannetti, S. Guha, S. Lloyd, L. Maccone, and J. H. Shapiro, 
Phys. Rev. A \textbf{70}, 032315 (2004); A. Serafini, J. Eisert, and
M. M. Wolf, Phys. Rev. A \textbf{71}, 012320 (2005).
\bibitem{macch-palma} C. Macchiavello and G. M. Palma, Phys. Rev. A \textbf{65},
050301(R) (2002).
\bibitem{mpv} C. Macchiavello, G.M. Palma and S. Virmani, 
Phys. Rev. A {\bf 69}, 010303(R) (2004).
\bibitem{PRA}
N.J. Cerf, J. Clavareau, C. Macchiavello and J. Roland,
quant-ph/0412089, to appear in Phys. Rev. A (2005).
\bibitem{fn1} In this work, we do not question the generally admitted
conjecture that Gaussian states achieve the classical capacity, so we
restrict our analysis to Gaussian input states.
\bibitem{fn2} Although the fraction
of the input mean photon number that is due to entanglement $\eta^*\to 0$ as $\bar{n}\to\infty$,
its absolute value $\eta^*\bar{n}$ tends to a constant (except for $x=1$).

\end{thebibliography}
\end{document}